\documentclass[12pt,preprint]{aastex}
\usepackage{epsfig}
\usepackage{rotating}
\usepackage{comment}
\usepackage{natbib}
\def\etal{{et al.\thinspace}}

\def\spose#1{\hbox to 0pt{#1\hss}}

\def\multleft#1{\hbox to size{\vbox {\halign {\lft{##}\cr #1}}\hfill}\par}
\def\multright#1{\hbox to size{\vbox {\halign {\rt{##}\cr #1}}\hfill}\par}

\def\degmark{^\circ}
\def\boxit#1{\vbox{\hrule\hbox{\vrule\kern3pt\vbox{\kern3pt
          #1 \kern3pt}\kern3pt\vrule}\hrule}}

\def\cm{{\rm\thinspace cm}}

\def\erg{{\rm\thinspace erg}}
\def\eV{{\rm\thinspace eV}}

\def\keV{{\rm\thinspace keV}}

\def\Msun{\hbox{$\rm\thinspace M_{\odot}$}}

\def\ph{{\rm\thinspace ph}}
\def\s{{\rm\thinspace s}}
\def\ks{{\rm\thinspace ks}}

\def\ergpcmsqps{\hbox{$\erg\cm^{-2}\s^{-1}\,$}}

\def\ergps{\hbox{$\erg\s^{-1}\,$}}

\def\pcmsq{\hbox{$\cm^{-2}\,$}}

\def\phpcmsqps{\hbox{$\ph\cm^{-2}\s^{-1}\,$}}

\makeatletter

\let\@internalcite\cite
\def\cite{\@ifstar{\citey}{\citefull}}
\def\citefull{\def\astroncite##1##2{##1\ ##2}\@internalcite}
\def\citey{\def\astroncite##1##2{##1\ (##2)}\@internalcite}
\def\citeyear{\def\astroncite##1##2{##2}\@internalcite}
\def\citename{\def\astroncite##1##2{##1}\@internalcite}
\def\@citex[#1]#2{\if@filesw\immediate\write\@auxout{\string\citation{#2}}\fi
  \def\@citea{}\@cite{\@for\@citeb:=#2\do
    {\@citea\def\@citea{; }\@ifundefined
       {b@\@citeb}{{\bf ??}\@warning
       {Citation `\@citeb' on page \thepage \space undefined}}%
{\csname b@\@citeb\endcsname}}}{#1}}

\def\@cite#1#2{#1\if@tempswa #2\fi}
\def\@biblabel#1{}

\def\astroncite#1#2{#1\ #2}
\makeatother

\begin{document}

\title{Measuring the Coronal Properties of
  IC~4329A with {\it NuSTAR}}

\author{L.~W.~Brenneman\altaffilmark{1}, 
G.~Madejski\altaffilmark{2},
F.~Fuerst\altaffilmark{3},
G.~Matt\altaffilmark{4},
M.~Elvis\altaffilmark{1},
F.~A.~Harrison\altaffilmark{3},
D.~R.~Ballantyne\altaffilmark{5},
S.~E.~Boggs\altaffilmark{6},
F.~E.~Christensen\altaffilmark{7},
W.~W.~Craig\altaffilmark{7,8},
A.~C.~Fabian\altaffilmark{9},
B.~W.~Grefenstette\altaffilmark{3},
C.~J.~Hailey\altaffilmark{10},
K.~K.~Madsen\altaffilmark{3},
A.~Marinucci\altaffilmark{4},
E.~Rivers\altaffilmark{3},
D.~Stern\altaffilmark{11},
D.~J.~Walton\altaffilmark{3},
W.~W.~Zhang\altaffilmark{12}}
\altaffiltext{1}{Harvard-Smithsonian CfA, 60 Garden St. MS-67,
  Cambridge, MA 02138, USA}
\altaffiltext{2}{Kavli Institute for Particle Astrophysics and
  Cosmology, SLAC National Accelerator Laboratory, Menlo Park, CA 94025, USA}
\altaffiltext{3}{Cahill Center for Astronomy and Astrophysics,
  California Institute of Technology, Pasadena, CA 91125, USA}
\altaffiltext{4}{Dipartimento di Matematica e Fisica, Universit\`{a} Roma
  Tre, via della Vasca Navale 84, I-00146 Roma, Italy}
\altaffiltext{5}{Center for Relativistic Astrophysics, School of
  Physics, Georgia Institute of Technology, Atlanta, GA 30332, USA}
\altaffiltext{6}{Space Science Laboratory, University of California, Berkeley,
  California 94720, USA}
\altaffiltext{7}{DTU Space—National Space Institute, Technical University of
  Denmark, Elektrovej 327, 2800 Lyngby, Denmark}
\altaffiltext{8}{Lawrence Livermore National Laboratory, Livermore, California
  94550, USA}
\altaffiltext{9}{Institute of Astronomy, Madingley Road, Cambridge CB3 0HA, UK}
\altaffiltext{10}{Columbia Astrophysics Laboratory, Columbia University, New
  York, New York 10027, USA}
\altaffiltext{11}{Jet Propulsion Laboratory, California Institute of
  Technology, Pasadena, CA 91109, USA} 
\altaffiltext{12}{NASA Goddard Space Flight Center, Greenbelt, Maryland 20771, USA}

\begin{abstract}
\noindent 
We present an analysis of a $\sim160 \ks$ {\it NuSTAR} observation of
the nearby bright Seyfert galaxy IC~4329A.
The high-quality broadband spectrum enables us to separate the effects
of distant reflection from the direct coronal continuum, 
and to therefore accurately measure the high-energy cutoff to be
$E_{\rm cut}=178^{+74}_{-40} \keV$.  
The coronal emission arises from accretion disk photons Compton
up-scattered by a thermal plasma, with the spectral index and cutoff being due
to a combination of the finite plasma temperature and optical depth.
Applying standard Comptonization models, we measure both physical
properties independently using the best signal-to-noise obtained to
date in an
AGN over the $3-79 \keV$ band.  We derive $kT_{\rm e}=37^{+7}_{-6} \keV$ with
$\tau=1.25^{+0.20}_{-0.10}$ assuming a slab geometry for the plasma,
and $kT_{\rm e}=33^{+6}_{-6} \keV$ with
$\tau=3.41^{+0.58}_{-0.38}$ for a spherical geometry, with both having
an equivalent goodness-of-fit.  
\end{abstract}  


\section{Introduction}
\label{sec:intro}

The primary hard X-ray continuum in Seyfert galaxies arises from
repeated Compton up-scattering of UV/soft X-ray
accretion disk photons in a hot, trans-relativistic plasma.  This process results in
a power-law spectrum extending to energies determined by the electron
temperature in the hot ``corona'' (for a
detailed discussion, see \citealt{Rybicki1979}).  The power-law index is a
function of the plasma temperature, $T$, and optical depth, $\tau$.  This scenario
describes the hard X-ray/soft $\gamma$-ray spectra of bright Seyferts
reasonably well (see, e.g., \citealt{Zdziarski2000}).  

There are few physical constraints on the nature of the corona.
Broad-band UV/X-ray spectra require that
it not fully cover the disk, and suggest it is probably patchy
\citep{Haardt1994}.  X-ray microlensing experiments suggest it is
compact; in some bright quasars a
half-light radius of $r_{\rm c} \leq 6\,r_{\rm g}$ has been measured
\citep{Chartas2009,Reis2013}, where we define $r_{\rm c}$ as the
radius of the corona and $r_{\rm g} \equiv GM/c^2$.  Eclipses of the
X-ray source have also placed constraints on the size of the hard X-ray
emitting region(s): $r_{\rm c} \leq 170\, r_{\rm g}$ (e.g.,
\citealt{Risaliti2007,Maiolino2010,Brenneman2013a}).

However, the coronal temperature and optical depth remain poorly
constrained due to the lack of high-quality X-ray 
measurements extending above $10 \keV$.   Complex spectral components,
including reflection from the accretion disk as well 
as from distant matter, contribute to the $\sim{\rm few}-30 \keV$ spectrum, and
constraints from previous observations suggest typical
cutoff energies are $E_{\rm cut} \geq 150 \keV$, requiring spectra
extending above $50 \keV$ for good constraints.
The {\em NuSTAR} high energy focusing X-ray telescope
\citep{Harrison2013}, which covers the band from $3-79 \keV$ with unprecedented
sensitivity, provides the capability to measure both the Compton
reflection component from neutral material and to constrain the continuum
cutoffs in bright systems.

The nearby Seyfert galaxy IC~4329A ($z=0.0161$, \citealt{Willmer1991};
Galactic $N_{\rm H}=4.61 \times 10^{20}
\pcmsq$, \citealt{Kalberla2005}; $M_{\rm BH}=1.20 \times 10^8 \Msun$,
\citealt{dlCP2010}) is a good candidate for
such measurements.  With a $2-10 \keV$ flux range
of $F_{\rm 2-10} \sim (0.1-1.8) \times 10^{-10}
\ergpcmsqps$ \citep{Beckmann2006,Verrecchia2007}, IC~4329A is one of the brightest
Seyferts.  In the hard X-ray/soft $\gamma$-ray band it appears similar
to an average radio-quiet Seyfert
(e.g., \citealt{Zdziarski1996}).   A Compton reflection component and strong
Fe K$\alpha$ line \citep{Piro1990} are both present.

IC~4329A has been observed in the hard X-ray band by {\it BeppoSAX}
\citep{Perola2002}, {\it ASCA+RXTE} \citep{Done2000} and {\it INTEGRAL} \citep{Molina2013}, which placed
rough constraints on the high-energy coronal cutoff at $E_{\rm cut}
\geq 180 \keV$, $E_{\rm cut}=150-390 \keV$ and $E_{\rm
  cut}=60-300 \keV$, respectively.  Combining {\it INTEGRAL} and {\it
  XMM-Newton} data further constrained the cutoff energy to $E_{\rm
  cut}=152^{+51}_{-32} \keV$ \citep{Molina2009}.  The soft
X-ray spectrum is absorbed by a combination of neutral and partially ionized
gas, with a total column of $\sim3 \times 10^{21} \pcmsq$
\citep{Zdziarski1994,Madejski1995,Steenbrugge2005},
comparable to the host galaxy's ISM column density
\citep{Wilson1979}.  After accounting
for the reflection component, the intrinsic spectral variability is modest
\citep{Madejski2001,Miyazawa2009,Markowitz2009}.

We report here on the {\it NuSTAR} portion of our simultaneous {\it
  NuSTAR+Suzaku} observation of IC~4329A.  The combined {\it NuSTAR+Suzaku} analysis
  will be described in a forthcoming paper (Brenneman \etal, in prep.).
In \S2, we report on the {\it NuSTAR} observations.  Our spectral analysis
follows in \S3, with a discussion of the coronal properties and their
implications in \S4.

\section{Observations and Data Reduction}
\label{sec:obs}

IC~4329A was observed by {\it NuSTAR} quasi-continuously from
August 12-16, 2012.  After eliminating Earth
occultations, passages through the South Atlantic Anomaly (SAA) and other periods
of high background, the {\it NuSTAR} observation
totaled $\sim162 \ks$ of on-source time for focal plane module A (FPMA) and $159 \ks$ for FPMB.  The
total counts after background subtraction in the $3-79 \keV$ band were
$426,274$ and $403,588$ for each
instrument, yielding signal-to-noise (S/N) ratios of $26.8$ and $24.6$, respectively.   

The {\it NuSTAR} data were collected with the FPMA and FPMB optical axes placed
roughly $\sim2$ arcmin from the
nucleus of IC~4329A.  We reduced the data using the {\it NuSTAR} Data
Analysis Software ({\sc nustardas}) and calibration version
1.1.1\footnote{http://heasarc.gsfc.nasa.gov/docs/nustar/}.  We filtered the event
files and applied the default depth
correction using the {\tt nupipeline} task.  We used circular extraction regions
$75$ arcsec in radius
for the source and background, with the source region centered on IC~4329A and the
background taken from the corner of the same detector, as close
as possible to the source without being contaminated by the PSF wings.
Spectra, images and light curves were extracted and response files were generated using the {\tt
  nuproducts} task.  
In order to minimize systematic effects, we have
not combined responses or spectra from FPMA and FPMB, but instead fit them
simultaneously.  We allow the absolute normalization for both modules to vary, and we find a
cross-calibration factor of $1.072 \pm 0.002$ for FPMB relative to FPMA.  

For all the analysis we used 
{\sc xspec} version 12.8.1, along with other {\sc ftools} packages within
HEASoft 6.14.  All errors in the text are 1$\sigma$ confidence within
the text unless otherwise specified, 
while the final parameter values and their uncertainties are quoted in Table~\ref{tab:tab1} at
$90\%$ confidence.  

IC~4329A demonstrated a modest, secular flux evolution during our observation,
increasing by $\sim12\%$ over the first $50 \ks$ of the
observation (using clock time), plateauing at maximum flux for $\sim50 \ks$
then decreasing by $\sim34\%$ over the remainder of the
observation.  On average, the flux we measure, $1.02 \times
10^{-10} \ergpcmsqps$, is within its historical range of values.  Given the lack of short
term variability and modest flux evolution we use the time-averaged spectrum
for all spectral analysis (\S\ref{sec:spectral}).

\section{Spectral Analysis}
\label{sec:spectral}

The focus of this paper is the characterization of the primary high-energy
continuum in IC~4329A, and we therefore
restrict our analysis to $5-79 \keV$ to avoid the effects of
low-energy absorption which are poorly constrained by
{\em NuSTAR} alone.   We assess the contributions of distant
reflection from the outer disk and/or torus, although we defer a detailed
discussion of the distant and inner disk reflection to a forthcoming
paper (Brenneman \etal, in prep.).

\subsection{Phenomenological Modeling of the {\em NuSTAR} Spectra}
\label{sec:phenom}

We first consider a phenomenological model for the {\it NuSTAR} spectra 
to generally investigate the prominence of the high-energy cutoff.  
We begin by evaluating the $5-79 \keV$  data against the {\tt
  pexmon} model \citep{Nandra2007}, which incorporates both
primary emission (in the form of a power-law with an exponential
cutoff) and distant reflection.  {\tt Pexmon}
is based on the {\tt pexrav} model of \citet{Zdziarski1995}, but in addition
to the reprocessed continuum emission it also includes the fluorescent emission
lines expected to accompany the Compton reflection (Fe K$\alpha$, Fe
K$\beta$, Ni K$\alpha$ and the Fe K$\alpha$
Compton shoulder).  

\begin{figure}%[H]
\centerline{
\includegraphics[width=0.4\textwidth,angle=270]{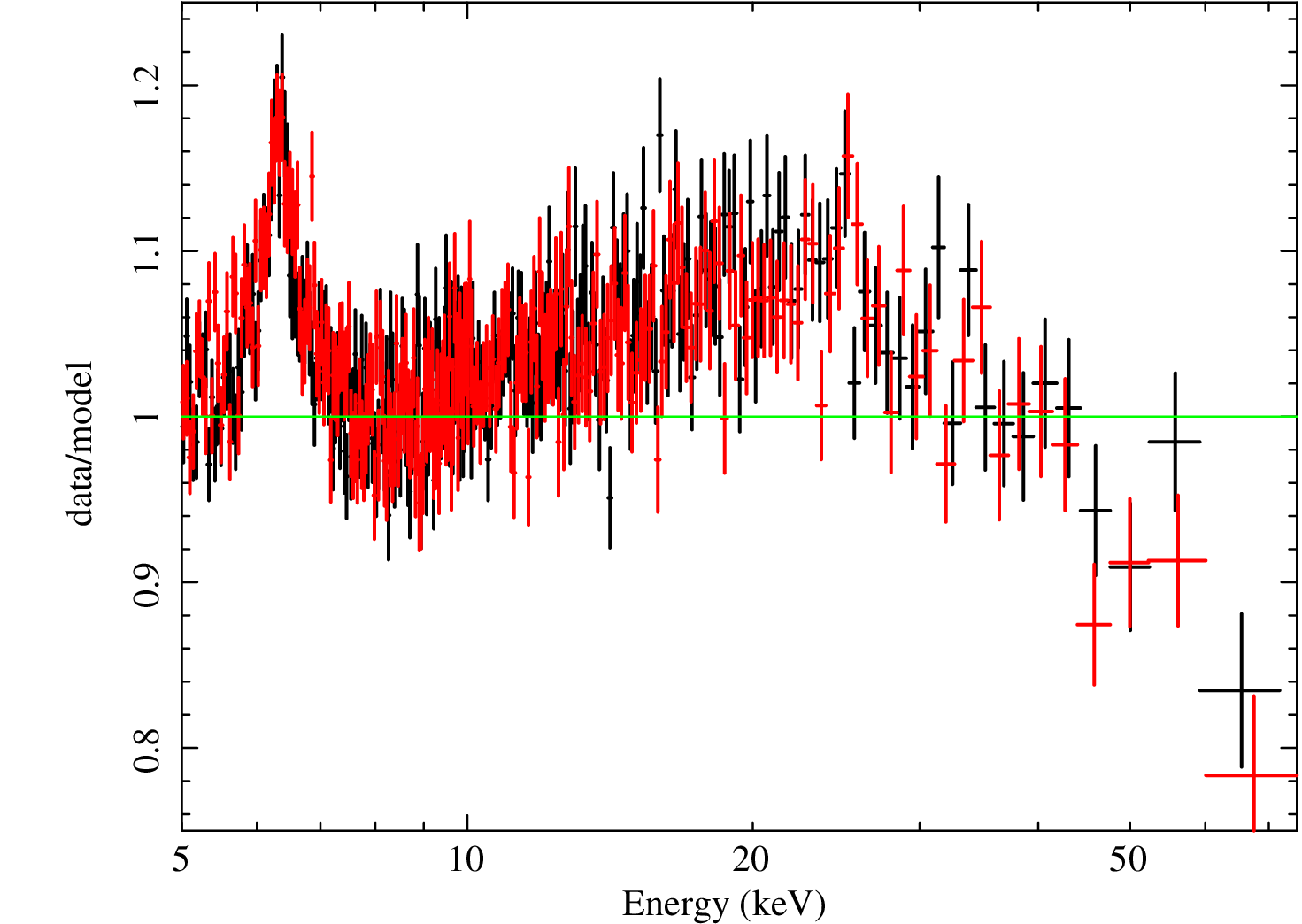}
\includegraphics[width=0.4\textwidth,angle=270]{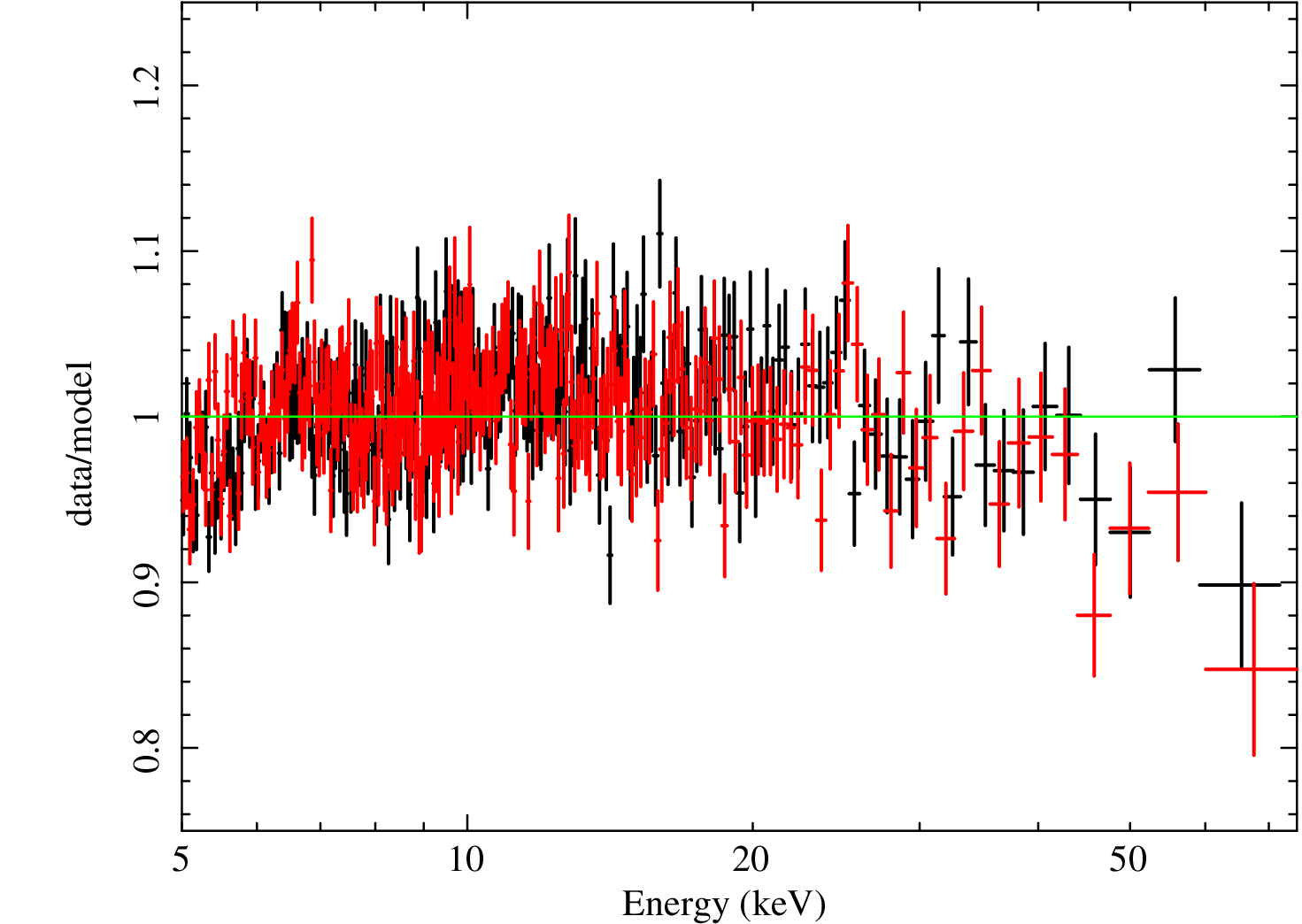}
}
\centerline{
\includegraphics[width=0.4\textwidth,angle=270]{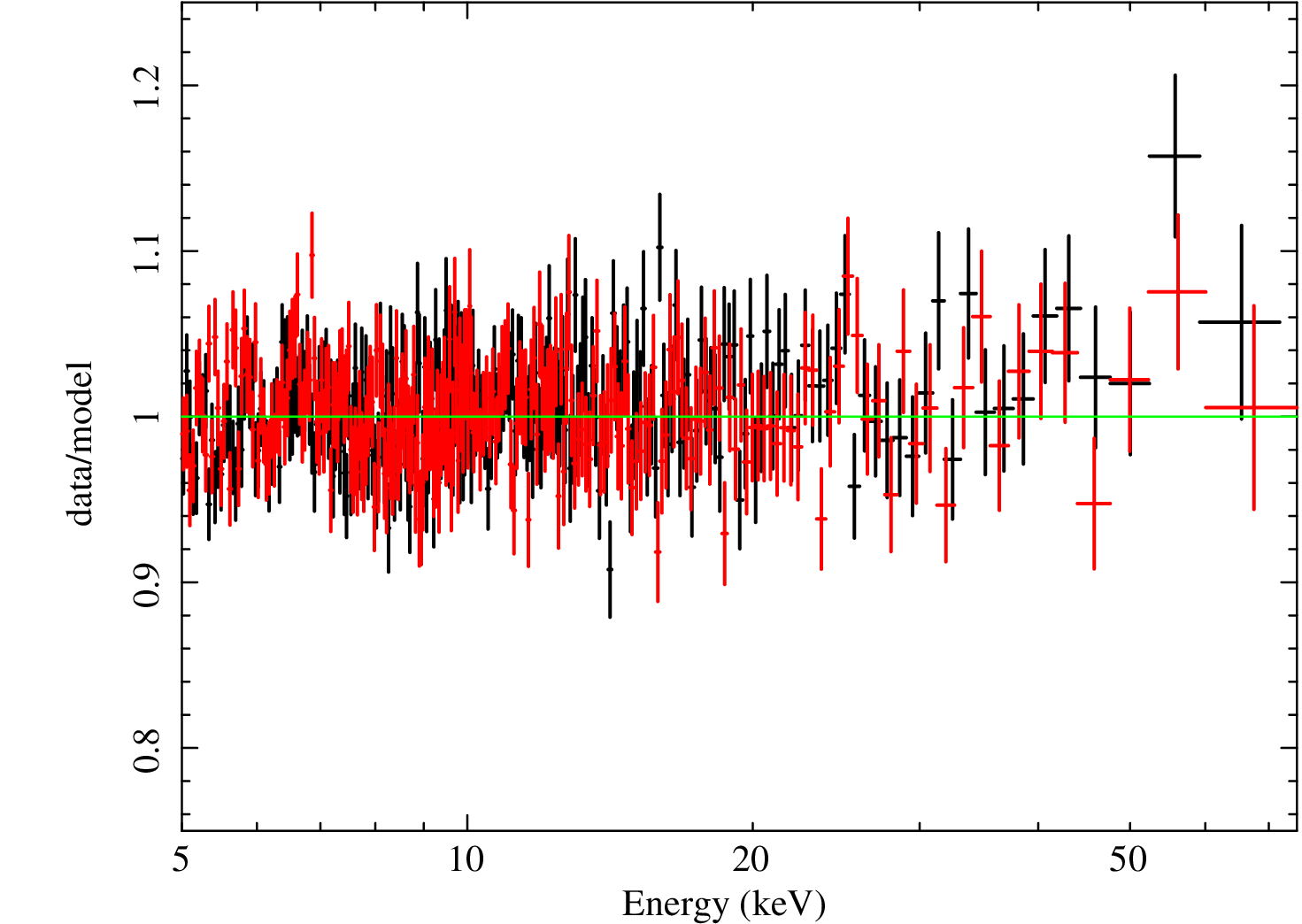}
\includegraphics[width=0.4\textwidth,angle=270]{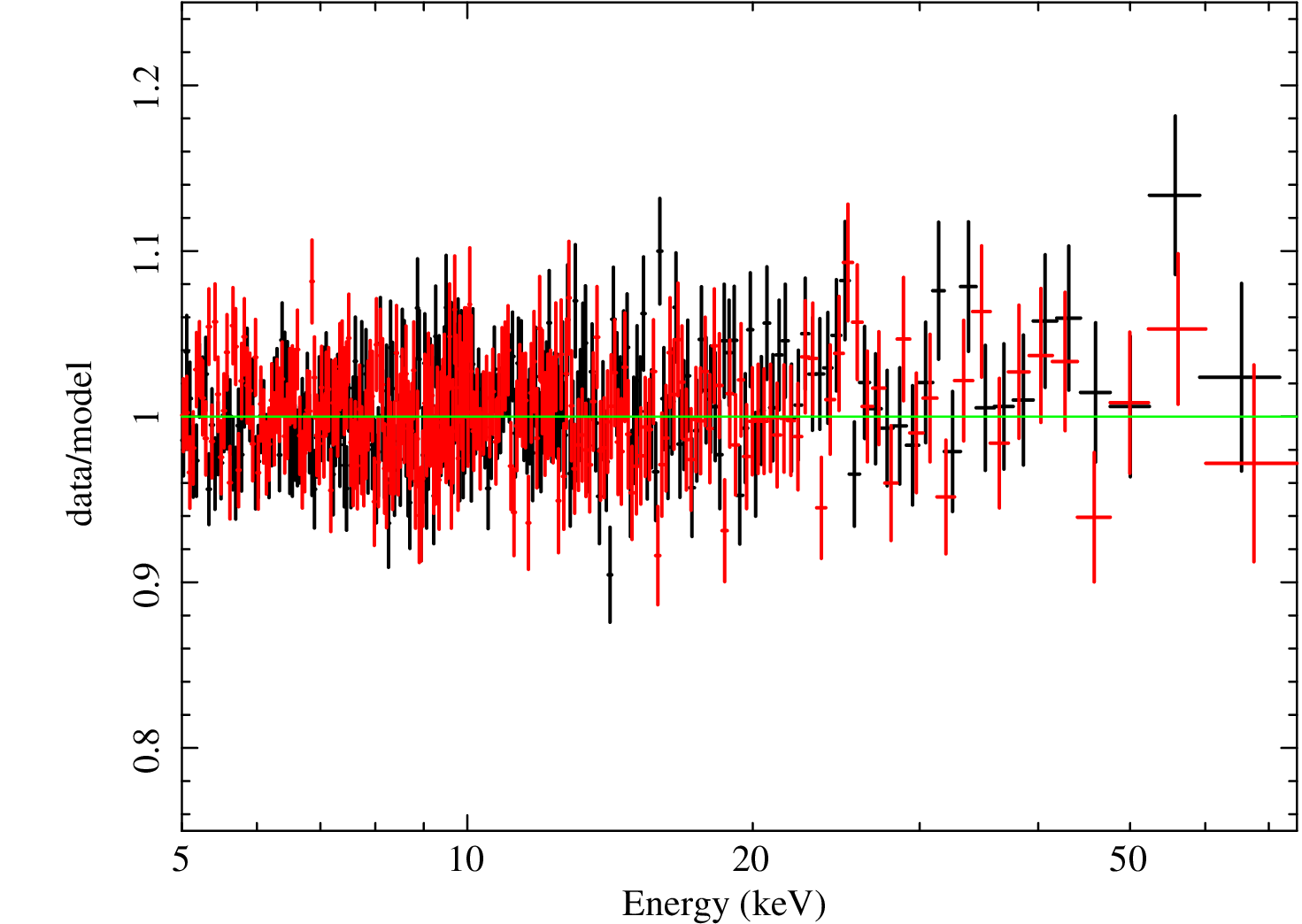}
}
\caption{{\small Plots of the data-to-model ratio of the {\it NuSTAR}/FPMA
    (black) and FPMB (red) spectra vs. a {\tt pexmon} model without an
    exponential cutoff or reflection (top left), including reflection
    (top right), including an exponential cutoff in addition to reflection
    (bottom left), and including also a broad Gaussian Fe K$\alpha$
    line (bottom right).  The horizontal green
    line in each plot represents a perfect fit.}}
\label{fig:fig1}
\end{figure}

Fitting the $5-79 \keV$ data using {\tt pexmon} with no exponential
cutoff or reflection ($E_{\rm cut}=10^6 \keV$ and $R=0$) yields a
goodness-of-fit of $\chi^2/\nu=2704/1723\,(1.57)$, with clear
residuals remaining in the Fe K band (Fig.~\ref{fig:fig1}, top).  The power-law has
a slope of $\Gamma=1.71 \pm 0.01$.  The spectrum has a pronounced
convex shape characteristic of
both Compton reflection and a high-energy cutoff above $\sim30
\keV$.  

Allowing the reflection component to fit freely, we get $R=-0.42 \pm 0.02$
(negative because of the way reflection is parameterized within the model; the
absolute value is the real reflection fraction)
with $\Gamma=1.83 \pm 0.02$, for
$\chi^2/\nu=2010/1722\,(1.17)$.  Clear 
residual curvature above $\sim25 \keV$ remains (Fig.~\ref{fig:fig1}, middle).  Also freeing the
cutoff energy of the primary
continuum yields $E_{\rm cut}=149 \pm 16 \keV$ with
$\Gamma=1.70 \pm 0.02$ and $R=-0.34 \pm 0.03$ with
$\chi^2/\nu=1881/1721\,(1.09)$ (Fig.~\ref{fig:fig1}, bottom).  It is
clear that both Compton reflection and a
high-energy cutoff are required: the improvement in fit of
$\Delta\chi^2/\Delta\nu=-129/-1$ ($\sim7\%$) upon addition of a high-energy cutoff is highly
significant.

Freeing the iron abundance (while
keeping the abundances of other elements fixed to their solar values)
improves the fit only slightly to $\chi^2/\nu=1871/1720\,(1.09)$, with
Fe/solar$=1.57 \pm 0.22$, though the uncertainties on the other
parameters increase slightly as a result.  Allowing the inclination
angle of the reflector to fit freely yields no constraints on the
parameter and no further improvement in fit, so we have elected to
keep it fixed at $i=60 \degmark$.  We note that when the {\tt pexmon} component
is replaced with the more common model of {\tt pexrav} plus Gaussian emission
lines, the fit yields similar values of $\Gamma$, $R$ and $E_{\rm cut}$.

Residuals still remain in the Fe K band, suggesting the presence
of an underlying broad component of the Fe K$\alpha$ line.  When this
feature is modeled with a Gaussian ($E=6.51 \pm 0.05 \keV$, $\sigma=0.36 \pm
0.03 \keV$, $EW=60 \pm 15 \eV$), the goodness-of-fit improves to
$\chi^2/\nu=1831/1717\,(1.07)$.  Including this component also lowers the iron
abundance to Fe/solar$=0.87 \pm 0.21$.  We refer to this model hereafter as
Model~1.  The modest broad iron line detection will
be discussed at length in Brenneman \etal (in prep.), and will not be further
addressed in this work.

\subsection{Toward a More Physical Model}
\label{sec:phys}

The high S/N of the data enable us to
consider more physically motivated models to describe the continuum emission.  
We employ the {\tt compPS} model of \citet{Poutanen1996}, which
produces the continuum through inverse Compton scattering of thermal disk photons off of
relativistic electrons situated above the disk in a choice of
geometries.  The disk photons and coronal plasma are each
parametrized by a single temperature.  It then reflects the
continuum X-rays off of a slab of gas (the disk or torus) to produce reprocessed
continuum emission self-consistently by internally linking with the
{\tt pexrav} model.  
Reprocessed line emission is not included.  

We fix the energy of the thermal disk photons to $kT_{\rm disk}=30 \eV$,
appropriate for a black hole of $M_{\rm
  BH} \approx 10^8 \Msun$ \citep{FKR2002}. We initially fix the geometry of the hot
electrons to be spherical with a
Maxwellian distribution, the former choice being based
on the coronal compactness measurements cited in \S1 (e.g., \citealt{Chartas2009}), 
although we also compare the results with a slab geometry.  We fit for
the electron energy ($kT_{\rm e}$) 
and coronal optical depth ($\tau$), as well as for the reflection
fraction ($R$) of the reprocessing gas, and model
normalization ($K$).  We add in separate Gaussian components to represent the
Fe K$\alpha$ (narrow and broad) and K$\beta$ line emission resulting from reflection in
order to maintain consistency with the {\tt pexmon} model employed in
\S\ref{sec:phenom} above.  

This approach yields approximately the same global goodness-of-fit as Model~1:
$\chi^2/\nu=1849/1715\,(1.07)$, with
parameter values of $kT_{\rm e}=48 \pm 22 \keV$ (assuming that $E_{\rm
  cut}=3kT_{\rm e}$, as per \citealt{Petrucci2001a}, this is equivalent
to a cutoff energy of approximately $144 \pm 66 \keV$, consistent with that
measured in Model~1), $\tau=2.70 \pm 1.31$
and $R=-0.39 \pm 0.06$, assuming 
a distant, neutral reflector inclined at $60 \degmark$ to the line of
sight.  We note that the iron abundance is not constrained by the
model.  The narrow Gaussian components representing narrow Fe K$\alpha$ and K$\beta$
%and broad Fe K$\alpha$ 
have equivalent widths of $EW_{\rm K\alpha}=41 \pm 8 \eV$ and $EW_{\rm
  K\beta}=15 \pm 12 \eV$, 
%and $EW_{\rm b}=49 \pm 15 \eV$, 
respectively.  The addition of these
components improves the global goodness-of-fit by $\Delta \chi^2 =
483$ and $\Delta \chi^2 = 19$, 
%and $\Delta \chi^2 = 12$, 
respectively, each for three additional degrees of freedom. 

We also note that consistent results are achieved (within
errors) when we fit for the Compton-y parameter rather than the
optical depth, as per the approach taken in \citet{Petrucci2013}:
$kT_{\rm e}=43 \pm 41 \keV$ and $y=1.16 \pm 0.05$,
resulting in $\tau=3.45 \pm 0.95$.  In addition to having larger
parameter uncertainties, however, this approach also results in an unconstrained
reflection fraction.  Consistent results are obtained when we
fix the reflection fraction at $R=-0.39$ (as described in the previous
paragraph) and Fe/solar$=1$: $kT_{\rm e}=39 \pm 9 \keV$, $y=1.06 \pm
0.01$ and $\tau=3.47 \pm 0.25$.  However, in light of the importance
of probing the reflection as a free parameter, we have elected to fit for the optical
depth explicitly in the model. 

We consider a slab geometry for the corona as well, again with a Maxwellian
electron distribution.  This results in no significant improvement to the fit:
$\chi^2/\nu=1836/1713\,(1.07)$, with $kT_{\rm e}=48 \pm 13 \keV$ (equivalent
to a cutoff energy of $E_{\rm cut}=144 \pm 39 \keV$, also consistent with that`
measured in Model~1), $\tau=1.50 \pm 0.39$ and $R=-0.38 \pm
0.07$.  This model is also insensitive to the Fe abundance.  The
equivalent widths of the Gaussian lines are the same as for the
spherical geometry, within errors.

For both geometries the electron plasma temperature is constant within errors, while
the optical depth is pushing the upper limit of its sensible parameter
space.  We therefore check the {\tt compPS} results using the 
{\tt compTT} model \citep{Titarchuk1994}, which assumes a simpler
thermal electron distribution and does not
self-consistently include the Compton reflection continuum.  We add
in reflection from distant matter using the {\tt pexmon} model.  We
assume an incident power-law photon index of $\Gamma=1.70$ (fixed),
as found in Model~1, and tie the normalization of the
reflected emission to that of the Comptonized component, such that the
contribution of the reflector relative to the Comptonization is
entirely determined by $R$ and the iron abundance.  We checked
the models to ensure a tight match between the shapes of a
power-law of this index and the shape of the {\tt compTT} component.  We also fix the
cutoff energy of the incident power-law at $E_{\rm cut}=3kT_{\rm e}$.  The spherical
geometry {\tt compTT} component model and the same model with a slab geometry
are referred to hereafter as Models~2-3, respectively.

We find values for the coronal temperature
consistent within errors with those of {\tt compPS} for both the
spherical and slab geometries: $kT_{\rm e}=33 \pm 11 \keV$ and $kT_{\rm e}=37 \pm 16
\keV$, respectively.  The
coronal optical depth in the slab geometry is slightly smaller with
{\tt compTT} vs. {\tt compPS}, though consistent
within errors: $\tau=1.25 \pm 0.50$.  For the spherical geometry, $\tau=3.41 \pm
0.90$, also consistent with its {\tt compPS} analog within errors.
The reflection fraction measured with this model (defined in the same
way for both
geometries) is comparable to that determined by {\tt compPS}: $R=-0.33
\pm 0.03$ for the sphere and $R=-0.32 \pm 0.04$ for the slab.  The iron
abundance is comparable that that found in
Model~1: Fe/solar$=0.63 \pm 0.20$ (sphere) and Fe/solar$=0.68 \pm
0.20$ (slab).

\begin{figure}%[H]
\centerline{
\includegraphics[width=0.5\textwidth,angle=270]{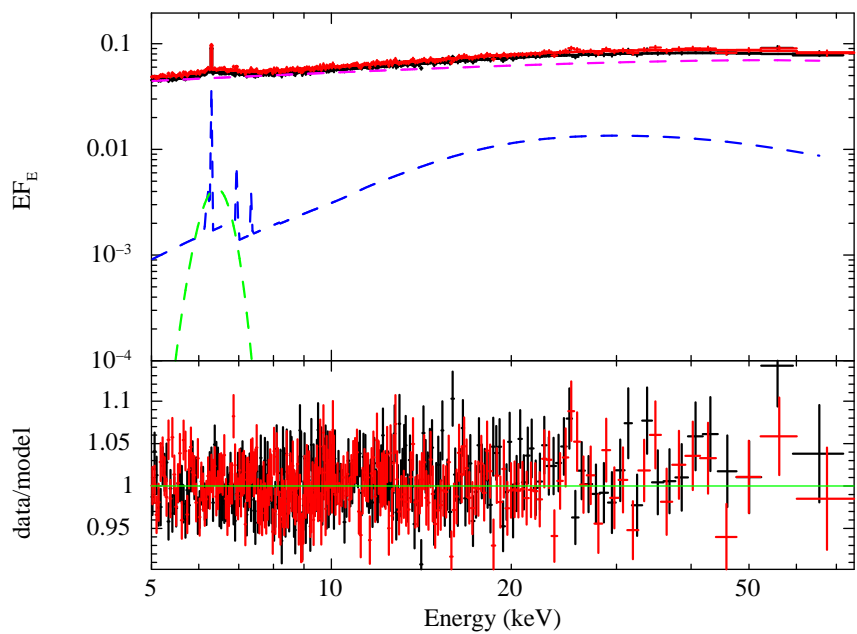}
}
\caption{{\small {\it Upper panel:} $EF_{\rm E}$ vs. energy plot for Model~1
    vs. the FPMA (black points) and FPMB (red points) data.  The {\tt pexmon} model
    incorporating the cutoff power-law continuum and distant reflection is in
    dashed red, the direct power-law emission is in dashed blue, and the broad
    Gaussian in dashed green.  The summed models are in solid black
    (FPMA) and solid red (FPMB).
    {\it Lower panel:} Data-to-model ratio for the FPMA (black) and
    FPMB (red).  The horizontal green line shows a ratio of unity for reference.}}
\label{fig:fig2}
\end{figure}

Table~\ref{tab:tab1} provides the best-fit parameters and their
Markov Chain Monte Carlo (MCMC)-derived
$90\%$ confidence errors for all three models.  Fig.~\ref{fig:fig2}
plots the contributions of individual
model components to the overall fit.  We show only the components for
Model~1, since
Models~2-3 look virtually identical, except using a Comptonization
component in lieu of a power-law.  The total absorbed $5-79
\keV$ flux and luminosity are $F_{5-79}=3.04 \times 10^{-10} \ergpcmsqps$
and $L_{5-79}=1.77 \times 10^{44} \ergps$, respectively.

\begin{figure}%[H]
\centerline{
\includegraphics[width=0.5\textwidth,angle=270]{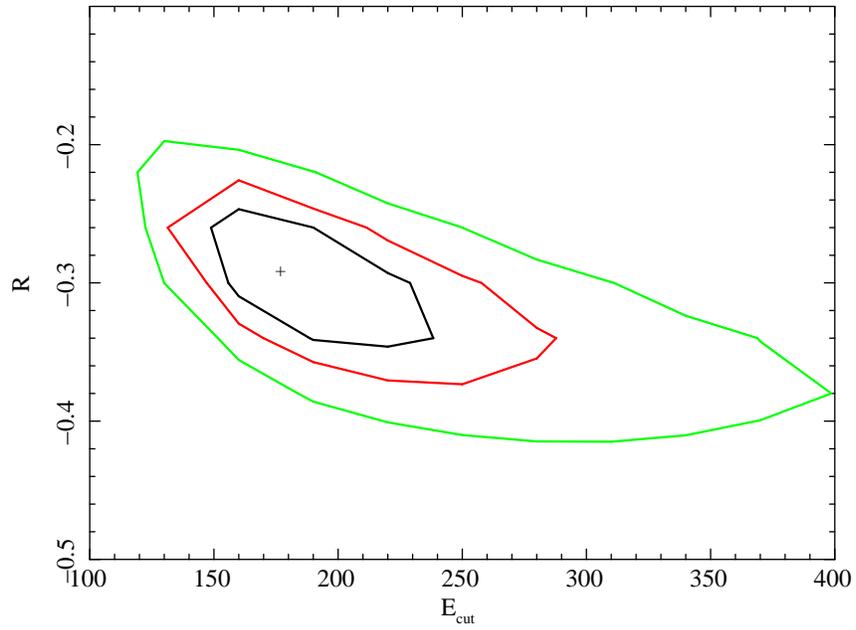}
}
\caption{{\small Output from the MCMC analysis of Model~1: contours
    show the $67\%$ (black), $90\%$ (red) and $99\%$ (green)
    probability densities for $R$ vs. $E_{\rm cut}$.}}
\label{fig:fig3}
\end{figure}

\begin{figure}%[H]
\centerline{
\includegraphics[width=0.5\textwidth,angle=270]{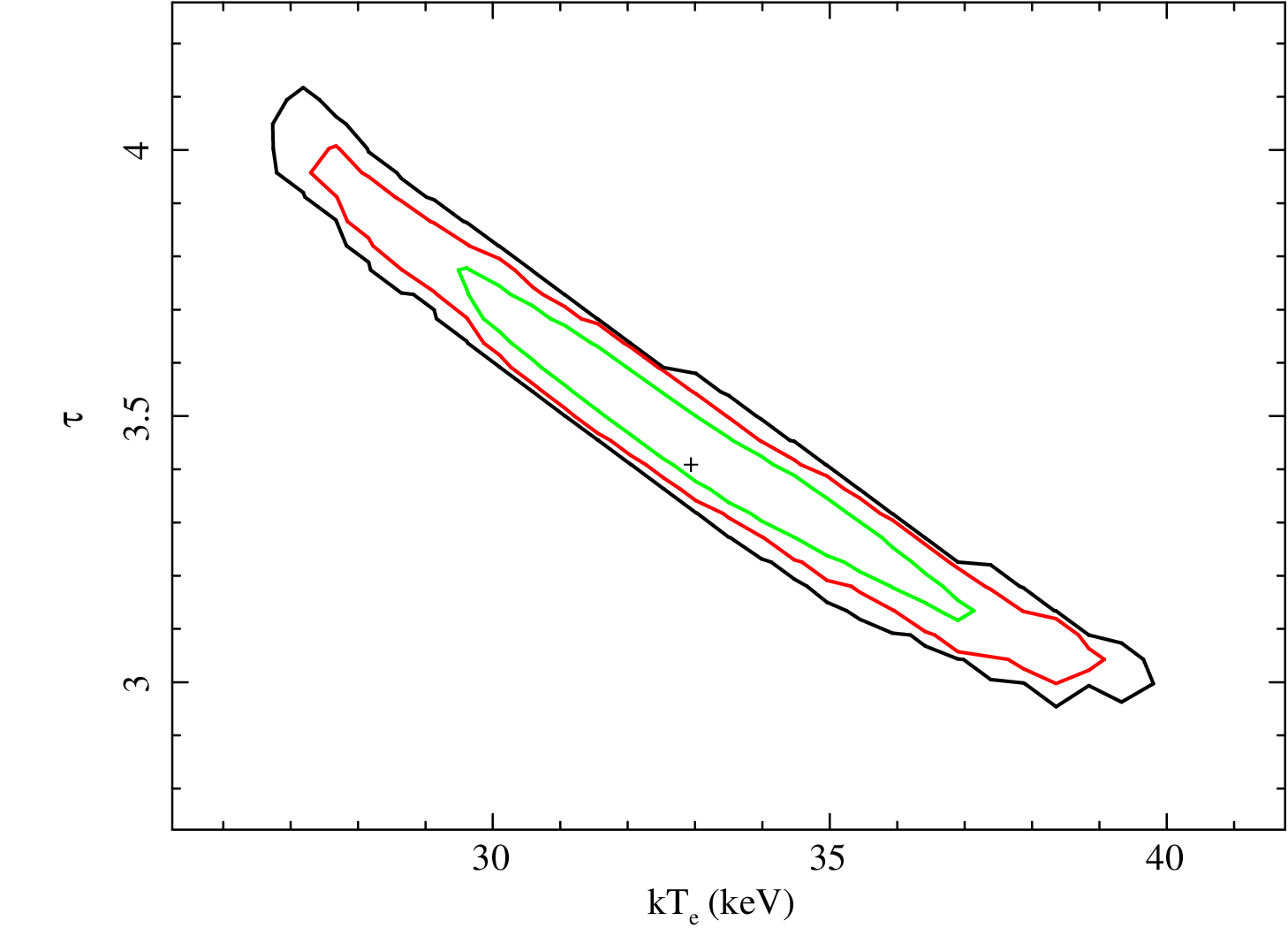}
}
\centerline{
\includegraphics[width=0.5\textwidth,angle=270]{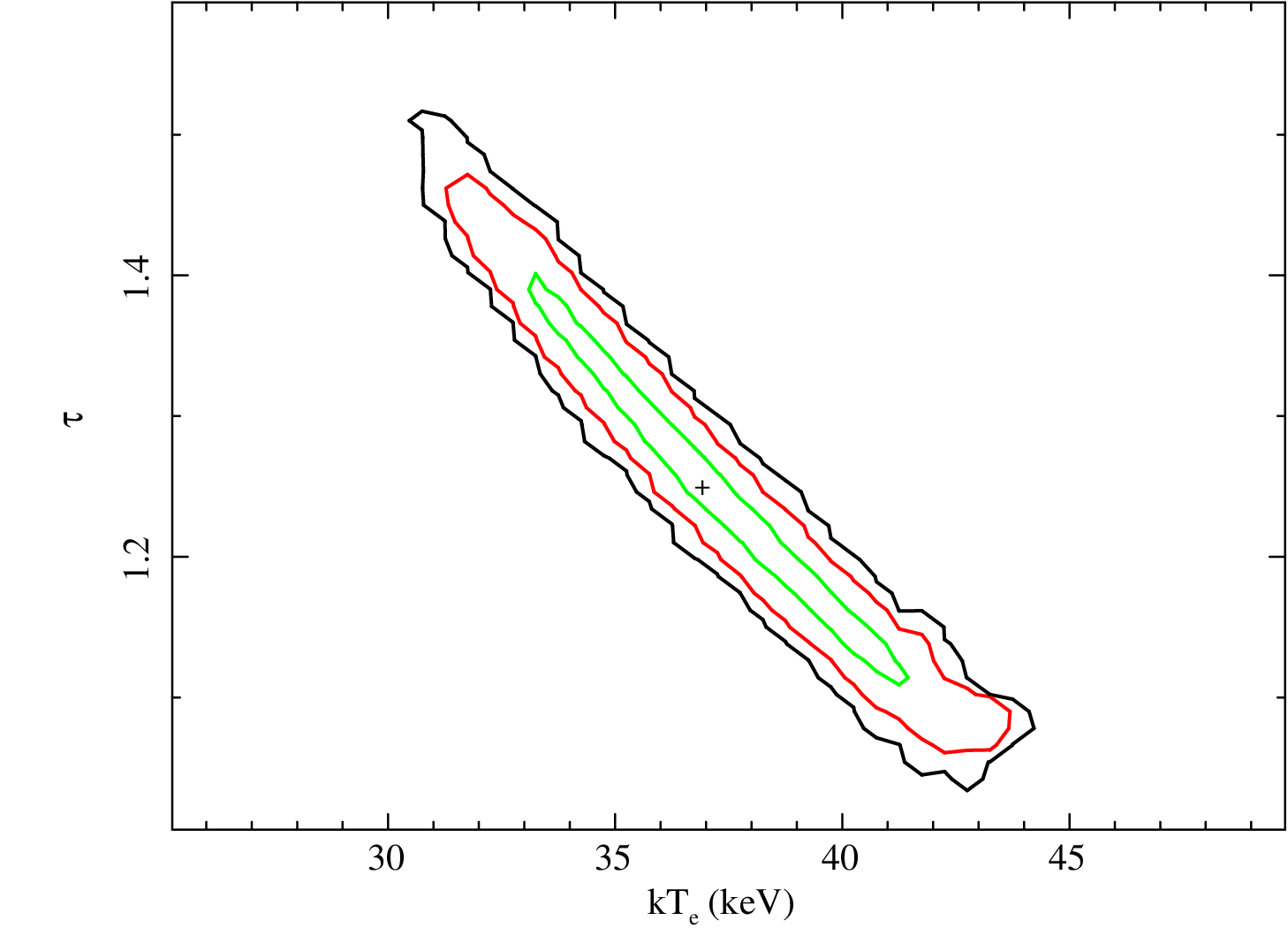}
}
\caption{{\small Output from the MCMC analysis of Model~2 (sphere; upper panel)
    and Model~3 (slab; lower panel):
    contours show the $67\%$ (green), $90\%$ (red) and $99\%$ (black)
    probability densities for $\tau$ vs. $kT_{\rm e}$.}}
\label{fig:fig4}
\end{figure}

The MCMC analysis employed to determine the formal parameter distribution used 
the Metropolis-Hastings algorithm (e.g., \citealt{Kashyap1998} and
references therein) 
following the basic procedures outlined in, e.g., \citet{Reynolds2012}.
Using this basic procedure we generated probability density contours for the most
interesting pairs of parameters for each model, shown in
Figs.~\ref{fig:fig3}-\ref{fig:fig4}.  Both sets of contours are
closed, implying that $kT_{\rm e}$ and $\tau$ are independently
constrained to $99\%$ confidence in both the spherical and slab
geometries.  Nonetheless, some degeneracy between the two parameters
still remains, as evidenced by the linear correlation seen in each
plot, due to an inherent modeling degeneracy between the
optical depth and temperature of the electron plasma in each geometry.
Fig.~\ref{fig:fig4} depicts the modest range over
which these parameters are degenerate, which is $\pm18\%$ of the
parameter space in
temperature and $^{+17\%}_{-11\%}$ in optical depth for the sphere, versus
$^{+19\%}_{-16\%}$ in temperature and $^{+16\%}_{-8\%}$ in optical
depth for the slab.

\begin{table*}%[H]
\begin{center}
%\hspace{-1.0cm}
\begin{tabular}{ccc@{}c@{}c}\hline\hline
{\bf Component} & {\bf Parameter (units)} & {\bf Model~1} & {\bf Model~2} & {\bf
  Model~3} \\
\hline \hline
{\tt TBabs} & $N_{\rm H}\,(\times10^{20} \pcmsq)$ & $4.61(f)$ & $4.61(f)$ & $4.61(f)$ \\
\hline
{\tt pexmon} & $\Gamma$ & $1.70^{+0.04}_{-0.04}$ & $1.70(f)$ & $1.70(f)$ \\
             & $E_{\rm cut}\,(\keV)$ & $178^{+74}_{-40}$ & $90^{+51}_{-18}*$ & $96^{+63}_{-24}*$ \\
             & $K_{\rm pex}\,(\times 10^{-2} \phpcmsqps)$ & $2.90^{+0.20}_{-0.18}$ & $2.90(f)$ & $2.90(f)$ \\
             & $R$ & $-0.29^{+0.06}_{-0.06}$ & $-0.33^{+0.04}_{-0.04}$ & $-0.32^{+0.04}_{-0.04}$ \\
             & Fe/solar & $0.87^{+0.42}_{-0.31}$ & $0.63^{+0.20}_{-0.20}$ & $0.68^{+0.21}_{-0.19}$  \\
\hline
{\tt compTT} & $kT_{\rm e}\,(\keV)$ & $---$ & $33^{+6}_{-6}*$ & $37^{+7}_{-6}*$ \\
             & $\tau$ & $---$ & $3.41^{+0.58}_{-0.38}$ & $1.25^{+0.20}_{-0.10}$ \\
             & $K_{\rm com}\,(\times 10^{-3} \phpcmsqps)$ & $---$ & $9.01^{+1.88}_{-2.95}$ & $8.17^{+2.37}_{-2.96}$ \\
\hline
{\tt zgauss} & $E_{\rm b}\,(\keV)$ & $6.51^{+0.05}_{-0.05}$ & $6.49^{+0.07}_{-0.06}$ & $6.48^{+0.08}_{-0.06}$  \\
             & $\sigma_{\rm b}\,(\keV)$ & $0.36^{+0.12}_{-0.10}$ & $0.36^{+0.09}_{-0.08}$ & $0.37^{+0.10}_{-0.08}$  \\
             & $K_{\rm b}\,(\times10^{-5} \phpcmsqps)$ & $7.84^{+2.72}_{-2.61}$ & $9.09^{+2.59}_{-2.50}$ & $9.11^{+2.64}_{-2.54}$ \\
             & $EW_{\rm b}\,(\eV)$ & $60^{+21}_{-20}$ & $70^{+20}_{-19}$ & $70^{+20}_{-20}$ \\
\hline \hline
Final fit & & $1831/1717\,(1.07)$ & $1836/1718\,(1.07)$ & $1837/1718\,(1.07)$  \\
\hline \hline
\end{tabular}
\hspace{-1.0cm}
\caption{\small{Best-fit parameters, their values and errors (to $90\%$
    confidence) for Models~1-3.  Parameters marked
    with an ({\it f}) are held fixed in the fit, while those marked with an (*) are
    tied to another parameter.}}
\label{tab:tab1}
\end{center}
\end{table*}

\section{Discussion}
\label{sec:disc}

The time-averaged $5-79 \keV$ {\it NuSTAR} spectrum is well-described
by a largely phenomenological
{\tt pexmon} model (Model~1) that includes, to $90\%$ confidence, a
continuum power-law with $\Gamma=1.70 \pm
0.03$, and a high energy cutoff  of $E_{\rm cut}=178^{+74}_{-40} \keV$, as well as
reprocessed emission from a distant
reflector ($R=-0.29^{+0.06}_{-0.06}$, Fe/solar$=0.87^{+0.42}_{-0.31}$).  Prior
measurements provided constraints on the cutoff energy of
the power-law at $E_{\rm cut} \geq 180 \keV$ \citep{Perola2002}, $E_{\rm
  cut}=150-390 \keV$ \citep{Done2000} and $E_{\rm
  cut}=60-300 \keV$ \citep{Molina2013}, though \citet{Molina2009} did achieve a
more precise constraint of $E_{\rm cut}=152^{+51}_{-32} \keV$ using {\it XMM} in
tandem with {\it INTEGRAL}.  The two datasets were not taken simultaneously,
however, and the S/N achieved by the {\it NuSTAR} data is superior to that of
{\it INTEGRAL}.  We therefore consider our new
measurements  ---which agree with all of the previously mentioned values,
within errors--- to be more robust.

The high-S/N, simultaneously obtained broad band {\em NuSTAR} spectrum enables us to 
apply physical models for the underlying coronal continuum emission
that go beyond phenomenological
descriptions.   The models 
parametrize the temperature, and optical depth of the electron
plasma for two coronal geometries; a sphere and a slab.  Both
geometries fit the data equally well, though we note that the sphere model
provides tighter constraints on its parameters.  Both models also produce
consistent values for the
electron temperature within errors.  However,
they result in slightly different
values for the optical depth:
$kT_{\rm e}=37^{+7}_{-6} \keV$ with
$\tau=1.25^{+0.20}_{-0.10}$ for the slab geometry, compared with $kT_{\rm e}=33^{+6}_{-6} \keV$ with
$\tau=3.41^{+0.58}_{-0.38}$ for the spherical geometry (both at $90\%$
confidence).  

This discrepancy  
in optical depth is due primarily to the way that the value is calculated for a
given geometry within the Comptonization models we employ (e.g.,
\citealt{Titarchuk1994,Poutanen1996}): the optical depth for a slab geometry is
taken vertically, whereas that for a sphere is taken radially and thus
incorporates an extra factor of $1/{\rm cos}(60)=2$.  Taking this
extra factor into account, the optical depth of the spherical case can
be translated into the slab geometry for ease of comparison:
$\tau=1.71^{+0.29}_{-0.19}$ for the sphere
vs. $\tau=1.25^{+0.20}_{-0.10}$ for the slab.  Though these values do
not formally agree within their $90\%$ confidence errors, they are
compatible at the $2\sigma$ level.

The derived electron temperatures for the sphere and slab coronal
geometries are low compared to $E_{\rm cut}/2$, but are not far off
from $E_{\rm cut}/3$, which is consistent with the corona having
significant optical depth (i.e., $\tau > 1$), modulo uncertainties in
geometry, seed photons, outflows, anisotropy, etc. which we are not
able to probe even with our high-S/N data.
Due to an inherent modeling degeneracy between the
optical depth and temperature of the electron plasma in each geometry,
there is a small, linearly correlated
range of values for these parameters which demonstrate approximately
equal statistical fit quality, as can be seen in
Figs.~\ref{fig:fig3}-\ref{fig:fig4}.  Nonetheless, we
constrain both parameters precisely and
accurately with the best data ever achieved over this energy band.
The data quality and goodness-of-fit of the Comptonization models gives us confidence in
the temperatures and optical depths we have measured.  We note, however, that
without high-S/N data at energies $\geq100 \keV$ we are unable to discriminate
between a thermal and non-thermal population of coronal electrons (e.g., with a
model such as {\tt eqpair}, \citealt{Coppi1999}).  A significant non-thermal
contribution could change the temperatures and optical depths that we measure.

With our robust determination of the continuum shape over the broad
energy range, we estimate that the power dissipated
in the corona, in the form of the power-law continuum, is $\sim87\%$ of the
total luminosity of the entire
system from $5-79 \keV$ (the power-law has a luminosity of
$L_{5-79}=1.53 \times 10^{44} \ergps$).
This represents $\sim10\%$ of the bolometric luminosity of
the source ($L_{\rm bol}= 1.60 \times 10^{45} \ergps$;
\citealt{dlCP2010}).  Given that IC~4329A is a radio-quiet
AGN (total 10 MHz - 100 GHz $L_{\rm r} = 2.2 \times 10^{39} \ergps$;
\citealt{Wilson1982}), we do not expect a significant portion of the X-rays to
come from a jet component, so we may infer that the remaining $\sim13\%$
of the $5-79 \keV$ emission represents the contribution from
reflection. 

Our spectral fitting results are broadly consistent with the signatures expected from
dynamic, outflowing coronae as defined by \citet{Beloborodov1999} and \citet{Malzac2001}: a
hard spectral index and relatively weak reflection.  If the corona is really
powered by compact magnetic flares that are dominated by $e^{\pm}$ pairs, the
resulting plasma is subject to radiation pressure from the photons moving
outwards from the disk.  The bulk velocity at which
the plasma should
be outflowing from the disk, given our Model~1 photon index of $\Gamma=1.70$, is
$v=0.22c$, according to \citet{Beloborodov1999}.  Our measured reflection fraction
of $R=0.3$ is also consistent with that predicted by \citet{Malzac2001} for a
Comptonizing plasma with these parameters.  We note that the
beaming of the coronal emission inherent in such dynamic models not
only hardens the spectrum, but also
implies that the intrinsic break in the spectrum occurs at lower
energies than what is observed: i.e., $E_{\rm cut} \sim 100 \keV$ rather than $E_{\rm
  cut}=178^{+74}_{-40} \keV$, as measured with Model~1.  A lower
intrinsic rollover is consistent with our measurements of the electron
temperature from Comptonization models ($kT \sim 35 \keV$), if we
assume that $E_{\rm cut}$ is between $2-3\,kT_{\rm e}$.

\acknowledgements{This work was supported under NASA Contract No. NNG08FD60C, and
made use of data from the {\it NuSTAR} mission, a project led by
the California Institute of Technology, managed by the Jet Propulsion
Laboratory, and funded by the National Aeronautics and Space
Administration.  We thank the {\it NuSTAR} Operations, Software and
Calibration teams for support with the execution and analysis of
these observations.  This research has made use of the {\it NuSTAR}
Data Analysis Software (NuSTARDAS) jointly developed by the ASI
Science Data Center (ASDC, Italy) and the California Institute of
Technology (USA).  LB gratefully
  acknowledges funding from NASA grant NNX13AE90G.  GM and AM acknowledge financial
  support from the Italian
Space Agency under contract ASI/INAF I/037/12/0 - 011/13.}

\bigskip

%% REFERENCES go in adsrefs.bib, get attached here

\bibliographystyle{apj}

\end{document}